\documentclass[shortnote,twocolumn]{jpsj3}
\usepackage{txfonts}
\usepackage{bm}
\usepackage{braket} 
\usepackage{graphicx}
\usepackage{color}
\usepackage{tikz}
\usepackage{quantikz}
\usepackage{float}
\usepackage{dcolumn}

\title{Low-Depth and Noise-Resilient Quantum State Preparation for Partial Differential Equations via Virtual $R_z$}

\author{Toru Fujii$^1$\thanks{Toru.Fujii@nikon.com}}
\inst{$^1$Nikon Corporation, 1-5-20, Nishioi, Shinagawa-ku, Tokyo 140-8601, Japan} 

\abst{
Preparing smooth real-amplitude states is essential in quantum PDE solvers such as LCHS, but exact decompositions are too deep for near-term devices. We propose a single-layer, hardware-aware ansatz with $R_y$ rotations, a CZ entangling layer, and virtual $R_z$ frame updates. By reducing physical pulses while preserving expressibility for smooth targets, it achieves substantially higher fidelity under realistic noise than deep exact constructions.
}

\begin{document}
\maketitle


The Linear Combination of Hamiltonian Simulation (LCHS) \cite{An23a,An23b} is a promising approach to dissipative partial differential equations (PDEs) \cite{Sato25}, where the target operator can be written in an integral form \(e^{-At} = \int_{\mathbb{R}} f(k) e^{-i(H + kL)t} \mathrm{d}k\) with strictly real and positive coefficients.
On noisy intermediate-scale quantum (NISQ) and early fault-tolerant (early-FTQC) architectures, a central bottleneck is the state-preparation oracle that encodes the square roots of these coefficients into amplitudes \cite{Ran20,Rudolph22,Berezutskii25}.
In typical PDE settings, the target amplitudes form a smooth, real-valued distribution, so exact universal state preparation is often unnecessarily costly.

Arbitrary state preparation can be implemented exactly via standard decompositions \cite{Mottonen04, Shende06, Plesch11}, but such methods synthesize generic complex unitaries in $U(N)$ and require circuit depths that scale exponentially with the number of qubits.
The resulting explosion in entangling gates causes severe fidelity loss under realistic noise.
This motivates a shallow, problem-tailored ansatz that exploits the smoothness and near-reality of the target amplitudes.

Here we propose a low-depth ansatz to prepare
$|\psi_{\mathrm{target}}\rangle=\sum_i \sqrt{w_i}\,|i\rangle$,
where $w_i$ are discretized weights arising from LCHS.
Figure~1 shows the single-layer circuit: local $R_Y$ rotations generate real amplitudes, CZ gates create limited entanglement, and additional $R_z$ rotations provide phase-frame flexibility.
Importantly, $R_z$ can be implemented virtually by updating the software phase reference of subsequent control pulses \cite{McKay17}.
Such virtual $R_z$ operations do not add physical pulses and therefore do not increase circuit duration; in our simulations we treat them as ideal frame updates (no additional noise and no additional time) to isolate the effect of providing extra $Z$-phase degrees of freedom without increasing the physical gate footprint.
From a tensor-network viewpoint, the alternating nearest-neighbor structure can be interpreted as restricting the state to a low-bond-dimension MPS \cite{Vidal03,Schollwock11} (with the required bond dimension bounded by the circuit depth across a cut; in our 3-qubit setting, $\chi\le 2$). This is well matched to smooth, weakly entangled distributions typical of damped PDE dynamics.

\begin{figure*}[b]
  \centering
  \includegraphics[width=140mm]{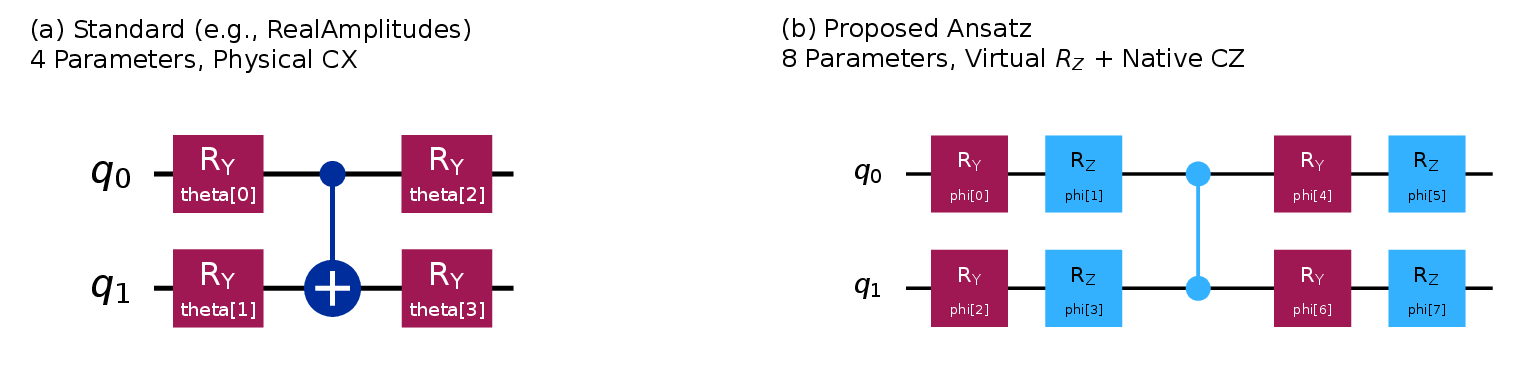}
  \caption{(Color online) Proposed single-layer ansatz (shown for $n=2$). The circuit uses $R_Y$ and $R_z$ rotations and a CZ entangling layer. The $R_z$ rotations (dashed boxes) are implemented as virtual phase-frame updates \cite{McKay17}, which do not add physical pulses or circuit duration; in simulations they are treated as ideal frame updates. The layer has $2n$ variational parameters.}
  \label{fig:MIS}
\end{figure*}

We benchmarked the proposed ansatz against (i) conventional exact state preparation compiled by Qiskit (optimization level 3) and (ii) the generic RealAmplitudes ansatz.
Table~I summarizes fidelities for $n=6$ and $n=12$ under two depolarizing-noise settings: a representative NISQ regime ($p_{2Q}=10^{-2}$, $p_{1Q}=10^{-3}$ plus readout errors) and an early-FTQC regime ($p_{2Q}=10^{-4}$, $p_{1Q}=10^{-5}$).
The exact approach achieves unit fidelity ideally but collapses under NISQ noise due to the exponential two-qubit depth (e.g., fidelity $0.0000$ at $n=12$).
In contrast, the proposed single-layer circuit attains high ideal fidelity already at depth $O(n)$ and maintains substantially higher fidelity under noise.
At $n=12$, the proposed 1-layer and RealAmplitudes(CZ) 1-layer circuits reach the same ideal fidelity ($0.9611$), indicating comparable expressibility at this depth, while deeper generic circuits (e.g., 2-layer RealAmplitudes) improve the ideal fidelity but suffer stronger noise accumulation in NISQ. In the noiseless setting, the ideal fidelity of the proposed ansatz does not improve from one to two layers, which is consistent with our real-amplitude targets: the additional virtual $R_z$ degrees of freedom become largely redundant and are optimized close to trivial phases. Under realistic noise—especially coherent errors—these virtual $R_z$ parameters instead provide effective knobs to absorb systematic phase/axis mismatches, leading to the marked robustness shown below.

\par\noindent\textit{Hardware/noise assumption.}
We target a device setting where $R_z$ rotations are implemented as \textit{virtual} (software-defined) phase updates (frame changes) \cite{McKay17} and thus contribute negligible physical gate error compared with driven rotations.
Motivated by CZ-native superconducting architectures, we model coherent calibration drifts primarily as (i) single-qubit $Z$-type phase over-rotations accompanying driven pulses and (ii) two-qubit $ZZ$-type coherent phase errors around entangling operations.
Accordingly, in the coherent-noise model below we apply a unitary $Z$-phase error to the \textit{physical} $R_Y$ gates and a unitary $ZZ$ phase error to the entangling gates, while assigning no additional noise or duration to virtual $R_z$ updates.

\par\noindent\textit{Optimization.}
 We optimized the variational parameters using the constrained optimization by linear approximation (COBYLA) algorithm by minimizing the infidelity \(1-F(\theta)\), where \(F(\theta)=|\langle \psi_t|\psi(\theta)\rangle|^2\). For each ansatz with \(P\) parameters, we set the maximum number of COBYLA iterations to \(\texttt{maxiter}=P\times m\) (with \(m=\texttt{max\_iter\_per\_param}\)), and used a stopping tolerance of \(\texttt{tol}=10^{-4}\). We performed \(S=\texttt{num\_seeds}\) multi-start runs with independent random initializations \(\theta_i\sim\mathcal{U}(-\pi,\pi)\), and report the best fidelity across seeds (and mean\(\pm\)std when stated).

To further assess robustness against coherent phase errors, we also performed a coherent-noise sweep in which a small $Z$ over-rotation is applied to one-qubit rotations and a stronger residual $ZZ$ phase error is applied to two-qubit gates (set to twice the one-qubit angle, as a representative parameterization).
To ensure a fair comparison with respect to the entangler, we use a CZ-entangler version of RealAmplitudes, denoted as RealAmplitudes(CZ), with the same CZ entangling pattern as the proposed circuit (i.e., we replace the default CX entangler by CZ while keeping the RealAmplitudes structure).
Using equal optimization budget proportional to the number of parameters (10 seeds), the proposed ansatz remains near unity fidelity across the sweep; for example at $(\theta_{1Q},\theta_{2Q})=(0.20,0.40)$ rad, the RealAmplitudes(CZ) baseline yields $0.9431\pm0.0181$ while the proposed ansatz yields $0.99992\pm0.00010$ (mean$\pm$std).

These results indicate that the proposed virtual-\(R_Z\)-enabled, low-depth circuit serves as a practical state-preparation primitive for PDE solvers on NISQ and early fault-tolerant hardware.

\begin{table}[htbp]
\centering
\caption{Fidelity comparison for $n=6$ and $n=12$.
“Qiskit Lvl3” denotes exact state preparation compiled with Qiskit optimization level~3.
“RealAmp(CZ)” is RealAmplitudes using a CZ entangler (the same CZ entangling pattern as the proposed circuit) for a fair comparison.
“1L/2L” are circuit layers.
Noise settings: NISQ ($p_{2Q}=10^{-2}$, $p_{1Q}=10^{-3}$ plus readout errors) and early-FTQC ($p_{2Q}=10^{-4}$, $p_{1Q}=10^{-5}$).}
\footnotesize
\setlength{\tabcolsep}{3.5pt}
\begin{tabular}{llrlrrr}
\hline\hline
Method & $n$ & Depth & Type & \multicolumn{3}{c}{Fidelity} \\
\cline{5-7}
 & & (2Q) & & Ideal & NISQ & e-FTQC \\
\hline
Qiskit Lvl3 & 6 & 57 & Exact & 1.0000 & 0.5295 & 0.9937 \\
Proposed 1L & 6 & 5 & Approx. & 0.9617 & 0.9036 & 0.9611 \\
RealAmp(CZ) 1L & 6 & 5 & Approx. & 0.9618 & 0.9037 & 0.9612 \\
Proposed 2L & 6 & 10 & Approx. & 0.9617 & 0.8542 & 0.9606 \\
RealAmp(CZ) 2L & 6 & 10 & Approx. & 0.9887 & 0.8782 & 0.9875 \\
Qiskit Lvl3 & 12 & 4094 & Exact & 1.0000 & 0.0000 & 0.6118 \\
Proposed 1L & 12 & 11 & Approx. & 0.9611 & 0.8401 & 0.9598 \\
RealAmp(CZ) 1L & 12 & 11 & Approx. & 0.9611 & 0.8401 & 0.9599 \\
Proposed 2L & 12 & 22 & Approx. & 0.9611 & 0.7432 & 0.9586 \\
RealAmp(CZ) 2L & 12 & 22 & Approx. & 0.9880 & 0.7640 & 0.9855 \\
\hline\hline
\end{tabular}
\end{table}



\acknowledgment
The author thanks Koshi Komuro, Kaito Tomari, Yosuke Okudaira, Yuichiro Hidaka and Shoichiro Tsutsui for useful discussions and programmings.







\end{document}